\documentclass[twocolumn,pra]{revtex4}
\usepackage{graphicx}
\usepackage{dcolumn}
\usepackage{amsmath}
\usepackage{bm}

\begin{document}

\title{Exotic dynamical evolution in a secant-pulse-driven quantum system}

\author{Peng-Ju Zhao}
\affiliation{Center of Theoretical Physics, College of Physical Science and
Technology, Sichuan University, Chengdu 610065, China}

\author{Wei Li}
\affiliation{Center of Theoretical Physics, College of Physical Science and
Technology, Sichuan University, Chengdu 610065, China}

\author{Hong Cao}
\affiliation{Center of Theoretical Physics, College of Physical Science and
Technology, Sichuan University, Chengdu 610065, China}

\author{Shao-Wu Yao}
\affiliation{Center of Theoretical Physics, College of Physical Science and
Technology, Sichuan University, Chengdu 610065, China}

\author{Li-Xiang Cen}
\email{lixiangcen@scu.edu.cn}
\affiliation{Center of Theoretical Physics, College of Physical Science and
Technology, Sichuan University, Chengdu 610065, China}

\begin{abstract}
We investigate an explicitly time-dependent quantum system driven by a
secant-pulse external field. By solving the Schr\"{o}dinger equation
exactly, we elucidate exotic properties of the system with respect to its
dynamical evolution: on the one hand, the system is shown to be essentially
nonadiabatic, which prohibits an adiabatic approximation for its dynamics;
on the other hand, the loop evolution of the model can induce a geometric
phase which, analogous to the Berry phase of the cyclic adiabatic evolution,
is in direct proportion to the solid angle subtended by the path of the state vector.
Moreover, we extend the model and show that the described properties coincide
in a special family of secant-pulse-driven models.
\end{abstract}

\maketitle

\section{Introduction}

Exact solution to driven quantum systems with time-dependent external fields
has long been a subject of particular interest in the field of quantum
mechanics \cite{qbook1,book1}. The motivation of the study comes not only
from the fundamental interest about the solvability of the driven quantum
system itself, but also from the nontrivial aspect of the dynamics that may
be generated through an explicitly time-dependent Hamiltonian. For example,
the cyclic adiabatic evolution of a time-dependent quantum system may induce
a geometric phase, the so-called Berry phase \cite{berry,simon}, which
indicates an intriguing connection between quantum physics and the gauge field
theory. Berry phase has been demonstrated to play important roles in various
areas of physics, e.g., in exploring the property of electrons
in crystals \cite{electron1,electron2} and in designing fault-tolerant
quantum operations for information processing \cite{HQC1,HQC2,HQC3,HQC4}.

In the light of the application to quantum control, the driven systems of
the following form have attracted considerable attention
\cite{LZ1,LZ2,wang,barnes,popul1,popul2,popul3,popul4,popul5}
\begin{equation}
H(t)=\Omega _xJ_x+\Omega _z(t)J_z,  \label{hamil1}
\end{equation}
where $J_{x,z}$ denote the angular-momentum operators satisfying
$[J_i,J_j]=i\varepsilon _{ijk}J_k$, the field component $\Omega_z(t)$
takes a time-dependent form and $\Omega_x$ is assumed to be a constant.
Relevant study on this type of driving protocols can be retrospected to the
original Landau-Zener model \cite{LZ1,LZ2}; the model and associated
variants have been extensively investigated and applied to, e.g., the
controllable quantum state transfer \cite{popul1,popul2,barnes,popul3,popul4,popul5},
Landau-Zener interferometry \cite{LZI1,LZI2,LZI3,LZI4}, and the charge
transfer and chemical reactions \cite{charge1,charge2,charge3,book1}. To our
best knowledge, the previous studies on this kind of driven systems had not
involved the geometric phase. In fact, a simple analysis on the adiabatic
solution of the Hamiltonian (\ref{hamil1}) displays that no geometric phase
could be generated in any such scanning protocols through the adiabatic
process. Let $|m\rangle $ denote the eigenstate of $J_z$ with magnetic quantum
number $m$. The instantaneous eigenstate of the Hamiltonian
is given by $|\psi _m^{ad}(\theta _{ad})\rangle=e^{i\theta _{ad}J_y}|m\rangle $ with
$\theta _{ad}=\arccos \frac{-\Omega_z}{\sqrt{\Omega _x^2+\Omega _z^2 }}$.
For any cyclic evolution the parameter $\theta_{ad}$ should retrace itself
and no Berry connection could be induced in the parametric space:
\begin{equation}
A_m\equiv i\langle \psi _m^{ad}(\theta _{ad})|\partial_{\theta _{ad}} |\psi
_m^{ad}(\theta _{ad})\rangle =-\langle m|J_y|m\rangle =0.  \label{gconnec}
\end{equation}
At this stage, we mention that this consequence resulted from the adiabatic
evolution and it does not indicate the necessity of restriction to the
nonadiabatic quantum process. So the question arises naturally: could one find a
system with the form of Eq. (\ref{hamil1}) that can generate
nonadiabatic geometric phase during its evolution?

The Aharonov-Anandan phase \cite{aaphase} has often been used to
describe the geometric phase for nonadiabatic processes.
Differing from the Berry phase in the cyclic adiabatic evolution,
the Aharonov-Anandan
phase can hardly be interpreted as a geometric object because it usually depends
on certain dynamical quantities, e.g., the rotating angular speed of the
evolving state vector \cite{aaphase1, aaphase2,xinqi}. In this article we
shall present a
driven model with the form of Eq. (\ref{hamil1}) and demonstrate that its
nonadiabatic evolution can generate the geometric phase. Astonishingly, we show
that the nonadiabatic geometric phase induced here is in close
analogy to the adiabatic Berry phase: a curvature vector could be identified
for the loop evolution in the Bloch space (instead of the parametric space);
the phase factor can then be understood as the geometric object of
the solid angle subtended by the closed path of the state vector. On the
other hand, our study reveals that the system is essentially nonadiabatic,
which prohibits an adiabatic approximation for its evolution.
So the existence of the geometric phase in the present model distinguishes
itself from the conventional Berry phase as it is rooted
in the nonadiabatic dynamics.

The rest of the article is organized as follows. In Sec. II we will
introduce the secant-pulse-driven model and solve the Schr\"{o}dinger
equation analytically by invoking a gauge transformation
approach. The explicit form of the dynamical invariant is achieved and the
solution of the system is then characterized in virtue of the
Lewis-Riesenfeld (L-R) theory \cite{lewis}. In Sec. IIIA we investigate the
nonadiabatic geometric phase induced by the loop evolution and elucidate
the linkage between it and the geometry of the evolving path in the
Bloch space. In Sec. IIIB we describe the essential characteristic of
the nonadiabaticity of the model. In Sec. IV we extend the system to a
more general form and show that the previously described properties
coincide in a family of secant-pulse-driven models. Finally,
a summary of the manuscript is presented in Sec. V.

\section{Exact solution to the secant-pulse-driven quantum model}

We address the driven model described by
\begin{equation}
H(t)=\hbar \nu (J_x-\frac 12\sec \frac{\nu t}2J_z),  \label{hamilsec}
\end{equation}
where the field component $\Omega _z(t)$ assumes a secant-shape pulse (see
Fig. 1) and the $x$ component $\Omega _x$ is fixed by $\Omega _x/\hbar =\nu $
(we set $\hbar =1$ afterwards) with $\nu $ the scanning frequency of $\Omega
_z(t)$. Consider the time evolution of the system during the pulsing
interval $t\in (t_0,t_f)$ with $|t_{0,f}|\leq \frac \pi \nu $. In view of
the Lie algebraic structure of the Hamiltonian, a potentially effective
way to solve the Schr\"{o}dinger equation
\begin{equation}
i\frac \partial {\partial t}|\psi (t)\rangle =H(t)|\psi (t)\rangle
\label{sequ}
\end{equation}
is to find out a specific gauge transformation \cite{alge}, $|\psi ^g(t)\rangle
=G^{\dagger }(t)|\psi (t)\rangle $, via which the system is
transformed into a new representation with a simpler Hamiltonian. Most of
the successful cases in previous studies \cite{popul3,popul4,popul5,wang}
have exploited the gauge transformation $G(t)$ in the form of
$e^{iz(t)J_z}e^{iy(t)J_y}$. Nevertheless, here we adopt a slightly
different form (although equivalent mathematically), $G(t)=e^{i\alpha
(t)J_x}e^{i\beta (t)J_y}$, and show that it is an efficient and convenient
choice for this particular model. Under this transformation, one obtains a
new Schr\"{o}dinger equation, $i\partial _t|\psi^g(t)\rangle =H^g(t)|\psi
^g(t)\rangle $, and the effective Hamiltonian $H^g(t)$ reads
\begin{eqnarray}
H^g(t) &=&G^{\dagger }(t)H(t)G(t)-iG^{\dagger }(t)\partial _tG(t)  \nonumber
\\
&=&\vec{X}(t)\cdot \vec{J},  \label{ghamil}
\end{eqnarray}
in which $X_i(t)$'s are given by
\begin{eqnarray}
X_1(t) &=&\dot{\alpha}\cos \beta -\frac \nu 2\sec \frac{\nu t}2\sin \beta
\cos \alpha +\nu \cos \beta ,  \nonumber \\
X_2(t) &=&\dot{\beta}+\frac \nu 2\sec \frac{\nu t}2\sin \alpha ,
\label{coeff} \\
X_3(t) &=&-\dot{\alpha}\sin \beta -\frac \nu 2\sec \frac{\nu t}2\cos \beta
\cos \alpha -\nu \sin \beta .  \nonumber
\end{eqnarray}
From Eqs. (\ref{coeff}), one can verify that by setting
\begin{equation}
\alpha (t)=\beta (t)=\frac 12(\pi -\nu t),  \label{angles}
\end{equation}
there are $X_{1,2}(t)=0$ and $X_3(t)=-\frac \nu 2\sec (\frac{\nu t}2)$, thus
\begin{equation}
H^g(t)=-\frac \nu 2\sec \frac{\nu t}2J_z.  \label{gham}
\end{equation}
As a result, the dynamical basis in this transformed representation is obtained
as $|\psi _m^g(t)\rangle =e^{-im\int_{t_0}^tX_3(\tau)d\tau}|m\rangle$ and
the basic solution to the original Schr\"{o}dinger equation (\ref{sequ}) is
then yielded via $|\psi _m(t)\rangle =G(t)|\psi _m^g(t)\rangle $. Not only
that, the above gauge transformation approach also indicates that the system
possesses a dynamical invariant, the so-called L-R invariant \cite{lewis},
\begin{eqnarray}
I(t) &=&G(t)J_zG^{\dagger }(t)  \nonumber \\
&=&-\sin \beta J_x+\cos \beta (\sin \alpha J_y+\cos \alpha J_z)  \nonumber \\
&=&-\cos \frac{\nu t}2J_x+\sin \frac{\nu t}2(\cos \frac{\nu t}2J_y+\sin
\frac{\nu t}2J_z),  \label{invar}
\end{eqnarray}
which satisfies $i\partial _tI(t)=[H(t),I(t)].$

\begin{figure}[t]
\includegraphics[width=0.88\columnwidth]{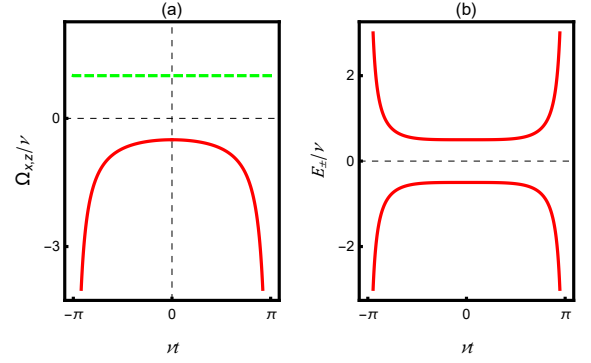}
\caption{The scanning process of the secant-pulse-driven model specified by
Eq. (\ref{hamilsec}): (a) The time dependency of the field component
$\Omega_z(t)/\nu$ during the interval $t\in(-\frac {\pi} {\nu},\frac {\pi} {\nu})$.
(b) The diabatic energy levels $E_\pm (t)$ over $\nu$ for the $j=\frac 12$ case.
The energy difference at $t=0$ is given by $E_+(0)-E_-(0)=-\nu$.}
\end{figure}

Let us express $I(t)$ as $I(t)\equiv \vec{R}(t)\cdot \vec{J}$ in which $\vec{R
}(t)=(\sin \theta \cos \varphi ,\sin \theta \sin \varphi ,\cos \theta )$ and
the angles $\theta (t)$ and $\varphi (t)$ are given by
\begin{equation}
\theta (t)=\arccos (\sin ^2\frac{\nu t}2),~~\varphi (t)=\pi -\arctan (\sin
\frac{\nu t}2).  \label{bloch}
\end{equation}
Note that these two equalities constitute the set of parametric equations for the
evolving trajectory of $I(t)$. In the parametric space spanned
by $\vec{R}(\theta ,\varphi )$, $I(t)$ will evolve along a fixed path $\chi $
on the surface of the unit sphere (see Fig. 2).
The orientation of $I(t)$ goes from $\theta =0$ at
$t_0=-\frac \pi \nu $ to $(\theta ,\varphi )=(\frac \pi 2,\pi )$ at
$t=0$, and then returns to the initial orientation at
$t_f=\frac \pi \nu $. According to the L-R theory, the
eigenvector $|\phi _m(t)\rangle $ of $I(t)$, specified by $I(t)|\phi
_m(t)\rangle =m|\phi _m(t)\rangle $, differs from the basic solution $|\psi
_m(t)\rangle $ of the system only by a phase factor: $|\psi _m(t)\rangle
=e^{i\Phi _m(t,t_0)}|\phi _m(t)\rangle $, where $\Phi _m(t,t_0)$ is given by
\begin{equation}
\Phi _m(t,t_0)=\int_{t_0}^t\langle \phi _m(\tau )|i\partial _\tau -H(\tau)
|\phi _m(\tau )\rangle d\tau .  \label{LRphase}
\end{equation}
The two terms of the above integration represent the geometric phase and
the dynamical phase, respectively. The kernel of the latter, accounting for
the diabatic energy levels of the system, is worked out to be
\begin{eqnarray}
E_m(t) &=&\langle \phi _m(t)|H(t)|\phi _m(t)\rangle   \nonumber \\
&=&-\frac{m\nu }2(\cos \frac{\nu t}2+\sec \frac{\nu t}2),  \label{level}
\end{eqnarray}
of which the two-level case, i.e., with the azimuthal quantum number $j=\frac 12$,
is depicted in Fig. 1(b).

\begin{figure}[t]
\includegraphics[width=0.85\columnwidth]{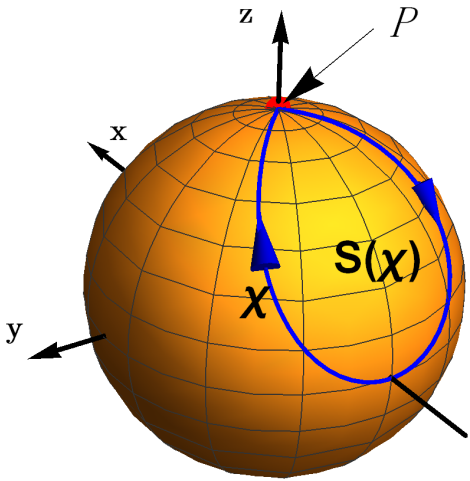}
\caption{Illustration of the loop evolution of the dynamical invariant $I(t)$
along the path $\chi$ in the parameter space spanned by $\vec{R}
(\theta,\varphi)$. It also characterizes the trajectory of the state vector
$|\phi_+(t)\rangle$ evolving on the surface of the Bloch sphere for the $j=\frac 12$
case. The enclosed surface of the loop evolution is denoted as $S(\chi)$.}
\end{figure}

\section{Exotic dynamical properties of the model}

\subsection{Nonadiabatic Berry phase induced in the model}

We now consider the geometric phase induced by the dynamical
evolution of the model. To be specific, we focus on the case of $j=\frac 12$.
The eigenstates of the dynamical invariant $I(t)$ in this case are
expressed as
\begin{eqnarray}
|\phi _{+}(t)\rangle  &=&\cos \frac{\theta (t)}2|+\rangle +\sin \frac{\theta
(t)}2e^{i\varphi (t)}|-\rangle ,  \nonumber \\
|\phi _{-}(t)\rangle  &=&\sin \frac{\theta (t)}2e^{-i\varphi (t)}|+\rangle
-\cos \frac{\theta (t)}2|-\rangle ,  \label{state2}
\end{eqnarray}
in which we have used the notation ``$\pm $'' for $m=\pm \frac 12$, respectively.
Up to a phase factor, the basis state $|\phi _{\pm }(t)\rangle $ undergoes a loop evolution
from the spin-up (spin-down) state $|\pm \rangle $ at $t\rightarrow -\frac \pi \nu $
to a state $\frac{\sqrt{2}}2(|+\rangle \mp |-\rangle )$ at $t=0$, and then
returns to the initial spin-up (spin-down) state at the ending point $
t\rightarrow +\frac \pi \nu $. The total phase $\Phi _{\pm }(t,t_0)$ induced
in the process can be written as
$\Phi _{\pm }(t,t_0)\equiv \Phi _{\pm }^d(t,t_0)+\Phi _{\pm }^g(t,t_0)$,
and the geometric phase $\Phi _{\pm }^g(t,t_0)$ is worked out to be
\begin{eqnarray}
\Phi _{\pm }^g(t,t_0) &=&\int_{t_0}^t\langle \phi _{\pm }(\tau )|i\partial
_\tau |\phi _{\pm }(\tau )\rangle d\tau   \nonumber \\
&=&\pm \int_{\nu t_0}^{\nu t}\frac{\cos ^3\frac q 2}{4(1+\sin ^2\frac q 2)}dq
\label{nadg}
\end{eqnarray}
with $q\equiv \nu \tau$. For the overall evolution with $t_{0,f}=\mp\frac \pi \nu $,
the above integration gives rise to $\Phi _{\pm }^g(t_f,t_0)=
\pm \frac 12(\pi-2)$.

To manifest the geometric feature of the above phase factor, let us change
the variable $t$ into $\vec{R}(\theta ,\varphi )$. In the Bloch space,
the basis state $|\phi _{+}(t)\rangle $ will evolve along the same path $\chi$
as that of $I(t)$ depicted in Fig. 2. The definite integral in the first line
of Eq. (\ref{nadg}) can then be recast as the line integral along the path
\begin{equation}
\Phi _{\pm }^g(t,t_0)=i\int_{\vec{R}_0}^{\vec{R}}\langle \phi _{\pm }(\vec{R}
)|\vec{\nabla}|\phi _{\pm }(\vec{R})\rangle \cdot d\vec{R}.  \label{nadg2}
\end{equation}
At this stage, it should be noted that the field component $\Omega _z(t)$
diverges as $t\rightarrow \pm \frac \pi \nu $ and the path of the state
vector is not strictly closed owing to the singularity at the point $\theta
=0$ (the point $P$ in Fig. 2). However, it is seen from
Eqs. (\ref{LRphase})-(\ref{nadg2}) that this divergency does not
occur in the loop integral of
the geometric phase $\Phi _{\pm }^g(\chi )$ but only affects that of the
dynamical part of the total phase $\Phi _{\pm }(\chi )$. So we can regard
the integral path of Eq. (\ref{nadg2}) a closed loop along which the
geometric phase can be calculated by integrating the curvature over the
enclosed surface
\begin{eqnarray}
\Phi _{\pm }^g(\chi ) &=&i\oint_\chi \langle \phi _{\pm }(\vec{R})|
\vec{\nabla}|\phi _{\pm }(\vec{R})\rangle \cdot d\vec{R}  \nonumber \\
&=&-\iint\nolimits_{S(\chi )}\vec{\nabla}\times \vec{\mathcal{A}}_{\pm }
(\vec{R})\cdot d\vec{S},  \label{Bphase}
\end{eqnarray}
where
\begin{equation}
\vec{\mathcal{A}}_{\pm }(\vec{R})\equiv i\langle \phi _{\pm }(\vec{R})|
\vec{\nabla}|\phi _{\pm }(\vec{R})\rangle  \label{gauge}
\end{equation}
denotes the nonadiabatic Berry connection. The occurring of the minus
in the second line of Eq. (\ref{Bphase}) is due to the clockwise direction
of the path $\chi$. It is direct to obtain
\begin{equation}
\vec{\mathcal{A}}_{\pm }(\vec{R})=
\mp \frac {\sin^2(\theta/2)} {R\sin\theta} \hat{\varphi}.
\label{curv}
\end{equation}
Thus the curvature $\vec{\nabla}\times \vec{\mathcal{A}}_{\pm }
(\vec{R})=\mp\frac {\hat{R}} {2R^2}$ and the surface integral in
Eq. (\ref{Bphase}) is obtained as
\begin{equation}
\Phi _{\pm }^g(\chi )=\pm \frac 12\iint\nolimits_{S(\chi )}\sin \theta
d\theta d\varphi =\pm \frac 12\Omega (\chi),  \label{sangle}
\end{equation}
where $\Omega (\chi)=\pi -2$ is just the solid angle swept by
the loop evolution. It is remarkable that the nonadiabatic geometric phase
induced here is independent of any dynamical quantity, e.g., the scanning
frequency $\nu$ or the angular speed of the evolving state vector.

In real physical systems the driving field cannot be infinite and
the truncation of the secant pulse is inevitable. The expression of
Eq. (\ref{nadg}) then describes the nonadiabatic geometric phase
for the non-cyclic dynamical process. It is worthy to note that,
as the integral kernel in Eq. (\ref{nadg}) tends to zero as
$\tau\rightarrow \pm\frac \pi \nu$, the truncation of the field pulse
results in very small influence on the amount of the geometric phase.
In Fig. 3 we depict the cutoff error to the geometric phase $\Phi_+^g(\chi)$
induced by the symmetric truncation
of the filed pulse, i.e., with $\nu t_{0,f}=\mp|\pi -\delta|$.
It is shown that the relative error is less than $10^{-3}$ even
when there is dramatic truncation $\delta\sim \frac {\pi} {10}$.

\begin{figure}[t]
\includegraphics[width=0.88\columnwidth]{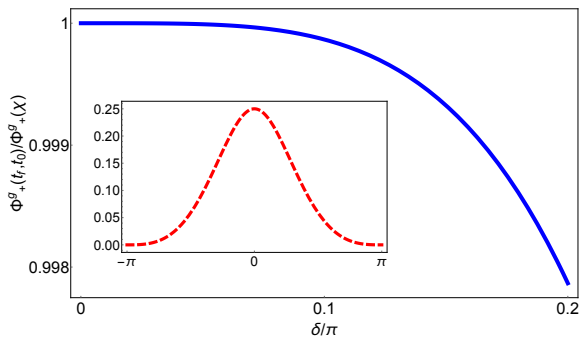}
\caption{Influence of the truncation of the field pulse
on the geometric phase. The curve of the inset describes
the integral kernel in
Eq. ({\ref{nadg}}) as a function of $q\equiv\nu\tau$.
It is shown that the relative error
[specified by $1-\Phi_+^g(t_f,t_0)/\Phi_+^g(\chi)$]
is almost negligible when the cutoff is within
$\delta\lesssim 0.2\pi $.}
\end{figure}

\subsection{Nonadiabaticity of the model}

In spite of the geometric feature described above, the nonadiabatic Berry
connection $\vec{\mathcal{A}}_{\pm }(\vec{R})$ obtained in Eq. (\ref{gauge})
could not recover the adiabatic one shown in Eq. (\ref{gconnec}). In particular,
we show in the below that the dynamical evolution generated by the present
model is essentially nonadiabatic and does not allow an adiabatic approximation.
That is to say, the geometric phase
induced in the model differs from the conventional Berry phase
since it is rooted in the nonadiabatic dynamics but not the commonly known
adiabatic process.

Firstly, we mention that the basis state $|\phi_m(t)\rangle$ of $I(t)$ could
never recover the adiabatic instantaneous eigenstate $|\psi_m^{ad}(t)\rangle$
of the Hamiltonian (\ref{hamilsec}). This can be intuitively understood in
view that in the parameter space $|\phi_m(t)\rangle$ always evolves along
the fixed trajectory which is independent of the scanning
frequency $\nu $ [cf. Eq. (\ref{bloch})]. More specifically, one can verify that the
quantitative condition of the adiabatic approximation,
\begin{equation}
\left| \frac{\langle \psi _{m}^{ad}(t)|\dot{\psi}_{n}^{ad}(t)\rangle}{
E_{m}^{ad}(t)-E_{n}^{ad}(t)}\right|= \left|\frac {\langle \psi _m^{ad}(t)|\dot{H}
(t)|\psi_n^{ad}(t)\rangle} {[E_m^{ad}(t)-E_n^{ad}(t)]^2}\right|\ll 1,  \label{adcon}
\end{equation}
cannot be fulfilled for the present model since both the numerator
and denominator above are dominated by the same power of the parameter $\nu$.
For the $j=\frac 12$ case a straightforward calculation gives
\begin{equation}
\left| \frac{\langle \psi _{+}^{ad}(t)|\dot{\psi}_{-}^{ad}(t)\rangle}{
E_{+}^{ad}(t)-E_{-}^{ad}(t)}\right|=\frac{|\sin \frac{\nu t}2|}2\sin \theta_{ad} \cos
^2\theta _{ad}  \label{adcon1}
\end{equation}
with $\theta_{ad}(t)=\arccos[(1+4\cos^2\frac {\nu t} {2})^{-1/2}]$. So
the adiabatic condition should be violated during the evolution
whatever how slow the scanning rate $\nu $ assumes.

Another perspective to exhibit the nonadiabaticity of the model
is to compare the state evolution generated by $H(t)$ of Eq. (\ref{hamilsec})
and that by $H^{\prime }(t)\equiv -H(t)$. Although the two Hamiltonians $H(t)$
and $H^{\prime }(t)$ possess completely identical instantaneous
eigenvectors, the dynamical evolution generated by them is different. In
view that $H^\prime (t)$ relates to $H(t)$ by a transformation
$\nu\rightarrow -\nu$, the dynamical invariant $I^\prime (t)$ (hence
the basis state $|\phi_m^\prime (t)\rangle$) of the model $H^\prime (t)$
can be obtained from $I(t)$ of the original
system $H(t)$ by changing $\varphi (t)$ into
$\varphi^\prime(t)=\pi +\arctan (\sin \frac{\nu t}2)$
but with $\theta^\prime(t)=\theta(t)$ [cf. Eq. (\ref{bloch})]. So the
nonadiabatic effect can be displayed by the loss of fidelity
between the basis sets of the two models:
$\delta_m(t)\equiv1-|\langle \phi _m(t)|\phi
_m^{\prime }(t)\rangle |^2$. For the $j=\frac 12$ case, one obtains
\begin{equation}
\delta_{\pm}(t)=\sin ^2\theta (t)\sin ^2\varphi (t)=\sin ^2\frac{\nu t}2\cos
^2\frac{\nu t}2.  \label{overlap}
\end{equation}
It is clear that the nonadiabaticity displayed above does not depends on the scanning
frequency $\nu $, which reconfirms that the model is essentially nonadiabatic.

\section{Generalization of the secant-pulse-driven model}

We now extend the above proposed model to a more general form of which
the field component $\Omega_x$ assumes an arbitrary
time-dependent scanning pulse. Specifically, this family of driven
models are shown as
\begin{equation}
\tilde{H}(t)=\Omega_x(t)\{J_x-\frac {1} {2}J_z\sec[\frac 12\int
\Omega_x(t)dt+\vartheta_0]\},
\end{equation}
in which $\Omega_x(t)$ is a general function of $t$ and the
constant $\vartheta_0$ in the integral $\vartheta(t)\equiv\frac 12\int
\Omega_x(t)dt+\vartheta_0$ is set such that $|\vartheta(t)|<\frac \pi 2$. By
invoking a similar gauge transformation $\tilde{G}(t)=e^{i\tilde{\alpha}(t)
J_x}e^{i\tilde{\beta}(t)J_y}$ with
\begin{equation}
\tilde{\alpha}(t)=\tilde{\beta}(t)=\frac {\pi} {2}-\vartheta(t),
\end{equation}
one can obtain an effective Hamiltonian in the transformed representation:
\begin{eqnarray}
\tilde{H}^g(t)&=&\tilde{G}^\dagger(t)\tilde{H}(t)\tilde{G}(t) -i\tilde{G}
^\dagger\partial_t\tilde{G}(t)  \nonumber \\
&=&-\frac {1} {2}\Omega_x(t)\sec \vartheta(t)J_z.  \label{hg2}
\end{eqnarray}
Subsequently the dynamical invariant of the system is achieved as
\begin{equation}
\tilde{I}(t)=-\cos\vartheta(t)J_x+\sin\vartheta(t)[\cos\vartheta(t)J_y+\sin
\vartheta(t)J_z].
\end{equation}
For the time interval during which $\vartheta(t)$ goes from $-\frac \pi 2$
to $\frac \pi 2$, $\tilde{I}(t)$ will evolve along the same path $\chi$ as
that of $I(t)$ shown in Fig. 2. So the geometric phase $\tilde{\Phi}_m(\chi)$
induced during the loop evolution is identical to $\Phi_m(\chi)$
[cf. Eq. (\ref{sangle})] obtained in the former model. That is to say, the
sort of driven models possess the universal geometric property with respect
to the dynamical evolution.

\section{Conclusion}

In summary, we have explored the dynamics generated by a secant-pulse-driven model.
The Schr\"{o}dinger equation of the system is solved
exactly by virtue of the gauge transformation approach. The nonadiabatic Berry phase,
or the so-called Aharonov-Anandan phase, induced by the loop evolution of
the model is shown to possess quite
exotic properties: it can be understood as the geometric object of the
solid angle subtended by the evolution path and is independent of the
evolving speed
of the state vector in the Bloch space; on the other hand, the geometric
phase achieved in the present model distinguishes itself from
the adiabatic Berry phase as the
model does not allow the adiabatic assumption for the
dynamical evolution. Furthermore, we have extended the system
to a more general form and show that the described feature
of the dynamics is universal in the specified family of secant-pulse-driven models.

For the potential application of the model, we note
that the spin geometric phase driven by magnetic field textures
has been exploited to manipulate electronic quantum states in
semiconducting nanostructures \cite{appl1,appl2}.
Very recently, the role of the nonadiabatic A-A phase of the
spin carriers subject to in-plane magnetic textures has also been
investigated in relation to the topological transition in electronic
spin transport \cite{appl3,appl4}. The model proposed in the present
manuscript offers a renewed way to address the relevant issue.
To this goal, a possible design of the described model in $1$D
conducting rings, which takes into account the matching of the
intrinsic Rashba field and the magnetic textures from an external source,
should be a research topic in the future study.


\begin{references}

\bibitem{qbook1} A. Bohm and M. Loewe, Quantum Mechanics: Foundations
and Applications (New York: Springer-Verlag, 3rd ed., 1993).


\bibitem{book1} H. Nakamura, Nonadiabatic transitions: Concepts, Basic
Theories and Applications (World Scientific, 2012).

\bibitem{berry} M.V. Berry, Proc. R. Soc. London A {\bf 392}, 45 (1984).
\bibitem{simon} B. Simon, Phys. Rev. Lett. {\bf 51}, 2167 (1983).

\bibitem{electron1} M.-C. Chang and Q. Niu, J. Phys. C {\bf 20},193202 (2008).
\bibitem{electron2} D. Xiao, M.-C. Chang, and Q. Niu, Rev. Mod. Phys. {\bf 82},
1959 (2010).

\bibitem{HQC1} P. Zanardi and M. Rasetti, Phys. Lett. A {\bf 264}, 94 (1999).
\bibitem{HQC2} J.A. Jones, V.Vedral, A. Ekert, and G. Castagnoli, Nature
(London) {\bf 403}, 869 (2000).
\bibitem{HQC3} L.M. Duan, J. I. Cirac, and P. Zoller, Science {\bf 292}, 1695
(2001).
\bibitem{HQC4} L.-X. Cen, X.Q. Li, Y.J. Yan, H.Z. Zheng, and S.J. Wang,
Phys. Rev. Lett. {\bf 90}, 147902 (2003).


\bibitem{LZ1} L.D. Landau, Phys. Z. Sowjetunion {\bf 2}, 46 (1932).
\bibitem{LZ2} C. Zener, Proc. R. Soc. A {\bf 137}, 696 (1932).

\bibitem{popul1} F.T. Hioe, Phys. Rev. A {\bf 30}, 2100 (1984).

\bibitem{wang} S.J. Wang and L.-X. Cen,  Phys. Rev. A {\bf 58}, 3328 (1998).

\bibitem{popul2} L.F. Wei, J.R. Johansson, L.X. Cen, S. Ashhab, Franco Nori,
Phys. Rev. Lett. {\bf 100}, 113601 (2008).

\bibitem{barnes} E. Barnes, S. Das Sarma, Phys. Rev. Lett. {\bf 109} 060401 (2012).



\bibitem{popul3} G. Yang, W. Li, L.-X. Cen, Chin. Phys. Lett.
{\bf 35}, 013201 (2018); arXiv: 1608.00735.
\bibitem{popul4} W. Li and L.-X. Cen, Ann. Phys. {\bf 389}, 1 (2018).
\bibitem{popul5} W. Li and L.-X. Cen, Quantum Inf. Process. {\bf 17}, 97 (2018).


\bibitem{LZI1} W.D. Oliver, Y. Yu, J.C. Lee, K.K. Berggren, L.S. Levitov,
T.P. Orlando, Science {\bf 310}, 1653 (2005).
\bibitem{LZI2} S.N. Shevchenko, S. Ashhab, F. Nori, Phys. Rep. {\bf 492}, 1 (2010).
\bibitem{LZI3} S. Gasparinetti, P. Solinas, J.P. Pekola, Phys. Rev. Lett. {\bf 107}, 207002 (2011).
\bibitem{LZI4} F. Forster, G. Petersen, S. Manus, P. H\"{a}nggi, D. Schuh,
W. Wegscheider, S. Kohler, S. Ludwig, Phys. Rev. Lett. {\bf 112}, 116803 (2014).




\bibitem{charge1} A.M. Kuztetsov, Charge Transfer in Physics, Chemistry, and Biology
(Gordon and Breach, Reading, 1995).

\bibitem{charge2} C. Zhu, S.H. Lin, J. Chem. Phys. {\bf 107}, 2859 (1997).
\bibitem{charge3} A. Nitzan, Chemical Dynamics in Condensed Phases
(Oxford University Press, 2006).



\bibitem{aaphase} Y. Aharonov and J. Anandan, Phys. Rev. Lett. {\bf 58}, 1593 (1987).

\bibitem{aaphase1} A. Bulgac, Phys. Rev. A {\bf 37}, 4084 (1988).

\bibitem{aaphase2} G.J. Ni, S.Q. Chen, and Y.L. Shen, Phys. Lett. A {\bf 197},
100 (1995).

\bibitem{xinqi} X.-Q. Li, L.-X. Cen, G. Huang, L. Ma, and Y.J. Yan,
Phys. Rev. A {\bf 66}, 042320 (2002).

\bibitem{lewis} H.R. Lewis Jr., Phys. Rev. Lett. {\bf 18}, 510 (1967).

\bibitem{alge} S.J. Wang, F.L. Li, and A. Weiguny, Phys. Lett. A {\bf 180} 189 (1993).

\bibitem{appl1} D. Loss, P. Goldbart, and A.V. Balatsky, Phys. Rev. Lett. {\bf 65}, 1655 (1990).

\bibitem{appl2} F. Nagasawa, D. Frustaglia, H. Saarikoski, K. Richter, and J.
Nitta, Nat. Commun. {\bf 4}, 2526 (2013).

\bibitem{appl3} H. Saarikoski, J.E. V\'{a}zquez-Lozano, J.P. Baltan\'{a}s,
F. Nagasawa, J. Nitta, and D. Frustaglia, Phys. Rev. B {\bf 91},
241406(R) (2015).

\bibitem{appl4} J.P. Baltan\'{a}s, H. Saarikoski, A.A. Reynoso, and D. Frustaglia,
Phys. Rev. B {\bf 96}, 035312 (2017).


\end{references}
\end{document}